\documentclass[prb,onecolumn]{revtex4}

\usepackage{amsmath}
\usepackage{graphicx}

\begin{document}

%%%%%%%%%%%%
\title{Double Point Contact in the $k=3$ Read-Rezayi State}

\author{Lukasz Fidkowski$^{1,2}$}
\affiliation{$^1$Microsoft Station Q\\
$^2$ Department of Physics, Stanford University, Stanford, CA 94305}

\date{\today }

\begin{abstract}
We compute the dependence of the tunneling current in a double point contact in the $k=3$ Read-Rezayi state (which is conjectured to describe an incompressible quantum hall fluid at filling fraction $\nu=12/5$) on voltage, separation between the two contacts, and temperature.  Using the tunneling hamiltonian of cond-mat/0607431, we show that the effect of quasiholes in the bulk region between the two contacts is simply an overall constant multiplying the interference term.  This is the same effect as found for the differential conductivity in cond-mat/0601242; the difference is that we do an actual edge theory calculation and compute the full current-voltage curve at weak tunneling.
\end{abstract}

\maketitle
%%%%%%%%%%%%
\section{Introduction}

Certain quantum systems in $2$ dimensions can support quasiparticles with anyonic and non-abelian statistics.  These are experimentally realized in fractional quantum hall (FQH) systems.  Specifically, the $\nu=5/2$ FQH state is conjectured to be in the universality class of the Moore-Read (MR) Pfaffian wavefunction, and it is believed that there is a FQH state at $\nu=12/5$ described by the $k=3, M=1$ Read-Rezayi (RR) state.  Both are non-abelian, and the latter supports universal quantum computation, making its study especially important.

In this paper we consider a double point contact geometry in an assumed $\nu=12/5$ RR state, and study tunneling in the edge theory.  A similar setup is considered in [\onlinecite{AK}].  The edge theory consists of a charged boson, and a neutral sector described by the ${\bf Z}_3$ parafermion CFT.  The latter contains spin fields $\sigma_i$ which are not local with respect to the other fields in the CFT, and it is precisely these that enter the hamiltonian describing tunneling of quasiholes, which is the most relevant tunneling process at low temperatures.  Unfortunately it is difficult to interpret such a tunneling Hamiltonian, because in a chiral theory, correlation functions of spin fields are not well defined.  Instead, they could be any linear combination of a finite number of {\it conformal blocks}.  In fact, it turns out that the choice of such linear combination is intimately related to the topology of the corresponding tunneling process occurring in the bulk; [\onlinecite{FFN}] give a prescription for determining the correct choice of linear combination, in the case of a single point contact.  We adapt their method to the case of a double point contact and use it to compute the tunneling current to first order in perturbation theory.  In contrast to the abelian case \cite{CFKSW}, we find that the interference term in the tunneling current is suppressed depending on the topological charge contained in the bulk region between the two point contacts.  A crucial point is that a bulk quasihole affecting the tunneling current must be prepared and annihilated on the edge in the far past and future respectively, and the prescription of [\onlinecite{FFN}] must be applied to this entire process.  After one divides out the universal factor associated with preparing and annihilating the bulk quasihole, one obtains correlation functions of the original spin fields which are finite and correctly incorporate the effect of the bulk quasihole.

\section{Setup}

We assume that there exists an incompressible quantum hall fluid at $\nu = 12/5$, and that it consists of a filled lowest Landau level (of both spins) and a first Landau level described by the $k=3, M=1$ Read-Rezayi state, which has filling factor $2/5$.  The Read-Rezayi wave function is a correlator of operators in the conformal field theory of a charged boson and ${\bf Z}_3$ parafermions.  The parafermions contain $6$ fields: $1, \sigma_1, \sigma_2, \psi_1, \psi_2, \epsilon$, and their fusion rules are determined by $\sigma_1 \sigma_2 = 1 + \epsilon$ and $\psi_1 \psi_2 = 1$ [\onlinecite{FFN}].  We will need the electron and quasihole operators; they are as follows \cite{EA}:

\begin{eqnarray}
\Phi_{\text{el.}} &=& \psi_1 e^{i \sqrt{\frac{5}{3}} \phi_c} \\
\Phi_{\text{q.h.}} &=& \sigma_1 e^{\frac{i}{\sqrt{15}} \phi_c}
\end{eqnarray} We have $q_{\text{el.}} = 1, \Delta_{\psi_1} = \frac{2}{3}, \Delta_{\text{el.}} = \frac{3}{2}$, and $q_{\text{q.h.}} = \frac{1}{5}, \Delta_{\sigma_1} = \frac{1}{15}, \Delta_{\text{q.h.}} = \frac{1}{10}$ .  The chiral edge theory also consists of ${\bf Z}_3$ parafermions and a charged mode, but they propagate with different velocities, $v_n$ and $v_c$.  Generally, one expects $v_c > v_n$.  In writing down the edge theory description of the tunneling hamiltonian, we a priori need to consider all sorts of tunneling processes, but a quick renormalization group argument \cite{FFN} shows that the most relevant term at low temperatures corresponds to an $e/5$-charge quasihole - the tunneling hamiltonian simply destroys such a quasihole on one side of the point contact and creates it on the other.  Here we assume we are at temperatures higher than those at which the crossover physics of [\onlinecite{FFN}], [\onlinecite{FLS}], etc. becomes important, so that the tunneling is weak and can be treated perturbatively.  This is an important point, because when we make reference to zero temperature results from now on we actually mean temperature small compared to $eV$, where $V$ is the bias voltage, and to $\hbar \omega_{\text{osc}}^c$ and $\hbar \omega_{\text{osc}}^n$.  Here $\omega_{\text{osc}}^c = 2 a / v_c$ and $\omega_{\text{osc}}^n = 2 a / v_n$ ($a$ being the distance between the two contacts) but still large compared to the crossover temperature.  Note that we have $4$ different energy scales, set by the temperature, bias voltage, $\omega_{\text{osc}}^c$ and $\omega_{\text{osc}}^n$.  The last two cannot be varied independently, but only through the variation of the separation $a$.  The two cases we consider in this paper are zero temperature, where effectively we have one dimensionless parameter we can vary (say $e V / (\hbar \omega_{\text{osc}}^c)$), and temperature high compared to $\hbar \omega_{\text{osc}}$, where things simplify as well.

The tunneling hamiltonian is

\begin{equation}
H_t(t) = \sum_j \Gamma_j e^{i \omega_J t} V_j (t) + \text{h. c.}
\end{equation} where $\Gamma_j$ is the tunneling amplitude at the $j$th point contact ($j = 1,2$), and $\omega_J = \frac{eV}{5 \hbar}$ is the Josephson frequency corresponding to a charge $e/5$ quasihole.  The tunneling operator $V_j$ is defined as

\begin{equation}
V_j (t) = \sigma_1 (x_{j, l}, t) \sigma_2 (x_{j,u}, t) e^{\frac{i}{\sqrt{15}} \phi_c (x_{j,l}, t)} e^{- \frac{i}{\sqrt{15}} \phi_c (x_{j,u}, t)}
\end{equation} where $x_{j,l}$ and $x_{j,u}$ are the spatial coordinates of the lower and upper sides of the $j$th point contact.  Note that because the hall bar is large but finite, the upper and lower points are actually on the same edge, just very far away from each other.  From now on it will be convenient to use the parametrization
\begin{eqnarray}
x_{j,l} = x_j \\
x_{j,u} = D - x_j
\end{eqnarray} where $D$ is the distance around the hall bar, which we will eventually take to infinity.

%An important point, and one which forms the crux of this paper, is that the operators $\sigma_i$ in the tunneling Hamiltonian are not well defined.  That is, there is a finite dimensional space of possibilities for an $n$-point correlator of such operators, called the space of {\it conformal blocks}.  What is really happening at the level of operators on Hilbert space is that the $\sigma_i$ are not well defined in certain sectors - there are multiple possibilities that satisfy the commutation relations required of $\sigma_i$, but no canonical choice.  This ambiguity leads to conformal blocks.  It turns out that what affects the choice of conformal block is determined by the topology of the tunneling process occurring in the bulk.  We use the prescription of [Fisher Fendley Nayak] to see how the topology of the tunneling processes in the bulk determines the choice of conformal block.  We will see that there is a unique, natural way to extend their prescription to the case of multiple point contacts with additional quasiparticles in the bulk, and that it gives a definite prediction for the tunneling current.

\section{Tunneling Current in Perturbation Theory}

Our discussion here closely parallels [\onlinecite{CFKSW}], except that we have the added complication of dealing with the parafermions and spin field operators.  As mentioned before, there is ambiguity in treating the $\sigma_i$ as operators, and their correlators are not well defined.  Indeed, to resolve the ambiguity we will need to add extra $\sigma_i$ operators, which create and annihilate the bulk quasiholes in the far past and future.  As we shall see, taking these to infinity essentially decouples them from the rest of the correlator, and their only influence is topological.  In fact, together with making the hall bar really large (that is, taking $D$ to infinity), the correlator of interest factors as a product of $2$-point functions, which is determined by conformal invariance up to a multiplicative constant.  All of the effects mentioned above are incorporated in that constant, denoted by $C_{jk}$ below.  In this section we calculate the tunneling current leaving $C_{jk}$ undetermined.  In the next section we determine $C_{jk}$.

We begin by writing down the appropriate operators.  The operator for the current across the point contact is the commutator of the total charge and the tunneling hamiltonian; because only the charge mode is involved the calculation is easy and we get

\begin{equation}
j(t) = \frac{ie}{5} \left( \sum_j \Gamma_j e^{-i \omega_J t} V_j(t) - \text{h. c.} \right)
\end{equation}To lowest order in perturbation theory, the expectation value of the tunneling current is

\begin{equation}
\label{eqn:int}
\langle j(t) \rangle = - i \int_{-\infty}^{t} dt' \langle 0 | [ j(t), H_t (t') ] | 0 \rangle.
\end{equation} We have

\begin{eqnarray}
\langle j (t) H_t (t') \rangle &=& \frac{ie}{5} \sum_{j,k} \langle \left( \Gamma_j e^{-i \omega_J t} V_j(t) - {\Gamma_j}^* e^{i \omega_J t} {V_j}^\dag (t) \right) \left( \Gamma_k e^{-i \omega_J t'} V_k (t') + {\Gamma_k}^* e^{i \omega_J t'} {V_k}^\dag (t') \right) \rangle \nonumber \\
&=& \frac{ie}{5} \sum_{j,k} \left( \Gamma_j {\Gamma_k}^* e^{-i \omega_J (t-t')} \langle V_j (t) {V_k}^\dag (t') \rangle - {\Gamma_j}^* \Gamma_k e^{i \omega_J (t-t')} \langle {V_j}^\dag (t) V_k (t') \rangle \right)
\end{eqnarray} because the only terms that contribute to the vacuum expectation value are ones that conserve the total charge on each side of the hall bar.  Now,

\begin{eqnarray}
\langle V_j (t) {V_k}^\dag (t') \rangle &=& 
\langle \sigma_1 (x_j,t) \sigma_2 (D-x_j,t) e^{\frac{i}{\sqrt{15}} \phi_c (x_j,t)} e^{-\frac{i}{\sqrt{15}} \phi_c (D-x_j,t)} \nonumber \\ & & \sigma_1 (D-x_k,t') \sigma_2 (x_k,t') e^{-\frac{i}{\sqrt{15}} \phi_c (x_k,t')} e^{\frac{i}{\sqrt{15}} \phi_c (D-x_k,t')} \rangle \\ & & \nonumber \\
&=& C_{jk} \langle \sigma_1(x_j,t) \sigma_2(x_k,t') \rangle \langle \sigma_2(D-x_j,t) \sigma_1(D-x_k,t') \rangle \nonumber \\ & & \langle e^{\frac{i}{\sqrt{15}} \phi_c (x_j,t)} e^{-\frac{i}{\sqrt{15}} \phi_c (x_k,t')} \rangle \langle e^{-\frac{i}{\sqrt{15}} \phi_c (D-x_j,t)} e^{\frac{i}{\sqrt{15}} \phi_c (D-x_k,t')} \rangle \\ & & \nonumber \\
&=& C_{jk} {\left[ \delta + i (v_n (t-t') + (x_j-x_k)) \right]}^{-2/15} {\left[  \delta + i (v_n (t-t') - (x_j-x_k)) \right]}^{-2/15} \nonumber \\ & & {\left[  \delta + i (v_c (t-t') + (x_j-x_k)) \right]}^{-1/15} {\left[  \delta + i (v_c (t-t') - (x_j-x_k))\right]}^{-1/15}.
\end{eqnarray} Here $C_{jk}$ is the constant mentioned above, which is determined by choice of conformal block (its phase is more subtle and convention dependent), and $\delta$ is an infinitesimal, specifying the choice of branch cut.  We will obtain all the information about $C_{jk}$ we need for our purposes in the next section.  For now, define
\begin{equation}
P(t,x) = \left[ \delta + i (v_n t + x) \right] ^{-2/15} \left[ \delta + i (v_n t - x)  \right] ^ {-2/15} \left[ \delta + i (v_c t + x) \right]^{-1/15} \left[ \delta + i (v_c t - x) \right]^{-1/15}.
\end{equation} so that

\begin{equation}
\langle V_j(t) {V_k}^\dag (t') \rangle = C_{jk} P(x_j-x_k, t-t').
\end{equation} In a similar manner we obtain

\begin{equation}
\langle {V_j}^\dag (t) V_k (t') \rangle = {C'}_{jk} P(x_j-x_k, t-t') = {C'}_{jk} P(x_k - x_j, -(t'-t))
\end{equation} Because ${V_j}^\dag (t)$ and $V_k (t)$ commute for spacelike separation, we have ${C'}_{jk} = C_{kj}$ (for $j=k$ the argument is a little more subtle but the conclusion holds - basically you have to infinitesimally spacelike separate the two operators and then argue using continuity).  We therefore have

\begin{eqnarray}
\langle [j(t), H_t (t')] \rangle &=& \frac{ie}{5} \sum_{j,k} ( \Gamma_j {\Gamma_k}^* e^{-i \omega_J (t-t')} C_{jk} [ P(x_j-x_k, t-t') - P(x_j-x_k, -(t-t')) ] \nonumber \\ & & - {\Gamma_j}^* \Gamma_k e^{i \omega_J (t-t')} C_{kj} [ P(x_k-x_j,t-t') - P(x_k-x_j,-(t-t')) ] )
\end{eqnarray}Doing the integral in (\ref{eqn:int}) we obtain

\begin{equation}
\langle j(t) \rangle = \frac{e}{10} \sum_{j,k} (C_{jk} \Gamma_j {\Gamma_k}^* + C_{kj} {\Gamma_j}^* \Gamma_k) [{\tilde P} (\omega_J, x_j-x_k) - {\tilde P} (- \omega_J, x_j-x_k)]
\end{equation} where ${\tilde P} (\omega_J, x)$ is the Fourier transform of $P(t,x)$.  For $x=0$ we can do the integral explicitly:

\begin{equation}
{\tilde P} (\omega, 0) = {v_n}^{-4/15} {v_c}^{-2/15} \frac{2 \pi}{\Gamma (2/5)} {| \omega |}^{-3/5} \Theta (\omega)
\end{equation} When $x \neq 0$, we can write down an alternate integral representation for ${\tilde P} (t,x)$ as follows: in the integral for the Fourier transform, the denominator is a product of $4$ terms, with $t$ (the variable being integrated over) appearing in each.  We replace the $t$ integration with an integration over $4$ variables $t_i$ ($i=1, \ldots, 4$), replacing $t$ with $t_i$ in each term in the denominator, and adding in delta function constraints to make all the $t_i$ equal.  This is just a rewriting of the original integral.  We then use the integral representation of the delta functions to obtain:

\begin{eqnarray}
{\tilde P} (\omega, x) &=& \frac {2 \pi}{{\Gamma(2/15)}^2 {\Gamma(1/15)}^2 {v_n}^{17/15} {v_c}^{16/15}} \int {d \omega_1}{d \omega_2}{d \omega_3} \exp \left[{ix \left( \frac{\omega_2 - 2 \omega_1 - \omega}{v_n} + \frac{\omega_2 - 2 \omega_3}{v_c} \right)}\right] \nonumber \\
& & {| \omega + \omega_1|}^{-13/15} {| \omega_2 - \omega_1 |}^{-13/15} {| \omega_3 - \omega_2 |}^{-14/15} {| \omega_3 |}^{-14/15} \Theta (\omega + \omega_1) \Theta (\omega_2 - \omega_1) \Theta (\omega_3 - \omega_2) \Theta (- \omega_3)
\end{eqnarray}Because the integrals over $\omega_j$ are all convergent in the above expression, we see that ${\tilde P}(\omega, x)$ is continuous as a function of $x$.  We write

\begin{equation}
{\tilde P}(\omega,x) = {\tilde P} (\omega, 0) H(\omega x / v_n, v_n/v_c)
\end{equation}With this notation we obtain
\begin{equation}
\langle j(t) \rangle = \frac{\pi e}{5 \Gamma (2/5)} {| \omega_J |}^{-3/5} \text{sgn} (\omega_J) {v_n}^{-4/15} {v_c}^{-2/15} \sum_{j,k} C_{jk} \Gamma_j {\Gamma_k}^* H(\omega_J (x_j - x_k) / v_n, v_c / v_n)
\end{equation}

\section{Finite temperature}

The analysis of finite temperature effects again proceeds much in the same way as in [\onlinecite{CFKSW}].  All the nontrivial topological information is still encoded in the $C_{jk}$.  The only difference between the zero temperature and finite temperature case arises from the fact that the $2$-point functions we now wish to use are analytic continuations of $2$-point functions on a cylinder.  The replacement that needs to be made is \cite{CFKSW}:

\begin{equation}
\frac{1}{{\left[ \delta + i (t \pm x / v_n) \right]}^{2/15}} \rightarrow {\left[ \frac{\pi T}{ \sin [ \pi T(\delta + i (t \pm x/v_n)) ]} \right]}^{2/15}
\end{equation}for the $\sigma_i$ correlator, and similarly for the correlator of the exponentials of the charge bosons, except with $v_n$ replaced by $v_c$.  We obtain the following modified $\tilde{P}$:

\begin{eqnarray}
{\tilde P} (\omega, x, T) = \int_{-\infty}^{\infty} (dt) e^{i \omega t} && {\left[ \frac{\sin [ \pi T(\delta+i(t+x/v_n))]}{\pi T} \right]}^{-2/15} {\left[ \frac{\sin [ \pi T(\delta+i(t-x/v_n))]}{\pi T} \right]}^{-2/15} \nonumber \\
&& {\left[ \frac{\sin [ \pi T(\delta+i(t+x/v_c))]}{\pi T} \right]}^{-1/15} {\left[ \frac{\sin [ \pi T(\delta+i(t-x/v_c))]}{\pi T} \right]}^{-1/15}
\end{eqnarray}

As opposed to the case of [\onlinecite{CFKSW}], this integral is not exactly doable for arbitrary values of $v_n, v_c$, the neutral and charge mode velocities.  However, we can extract the behavior of the current at large values of the temperature.  To do this, simply note that the integrand above is basically a product of inverse powers of $\sinh$, which at $T$ large compared to $\hbar \omega_{\text{osc}}^c$ and $\hbar \omega_{\text{osc}}^n$ become decaying exponentials, suppressed everywhere except at $t= \pm x/v_n, \pm x/v_c$.  Noting that the integral is everywhere convergent, we see that most of its contributions come from those $4$ points.  Plugging in, we see that we have an exponential suppression of

\begin{equation}
\exp \left[- 2 \pi T |x_1-x_2| \left( \frac{2/15}{v_n} + \frac{1/15}{v_c} \right) \right]
\end{equation} of the interference term at high temperatures $T \gg \hbar \omega_{\text{osc}}^c, \hbar \omega_{\text{osc}}^n$.  Comparing this to the result of [\onlinecite{CFKSW}] who obtain

\begin{equation}
\exp \left[ -2 \pi g T |x_1 - x_2| \right]
\end{equation} we see that we have an ``effective" $g$ of $2/(15 v_n) + 1 / (15 v_c)$.

\section{Determining the $C_{jk}$}

We now come to the crux of the problem, which is to find the constants $C_{jk}$.  This basically comes down to determining the conformal blocks for the $\sigma_j$ correlators, and here the topology comes into play.  Working in the context of the $\nu = 5/2$ state and a single point contact, [\onlinecite{FFN}] argued that for a given quasihole tunneling process, the two $\sigma$ fields that create and destroy the quasihole should fuse to the identity.  Arguing along these lines, one can uniquely determine the desired conformal block.  It is here that one uses the connection between the bulk and the boundary: the tunneling topology in the bulk determines the fusion channels of the tunneling operators in the boundary field theory.  To give an actual argument for this connection would require a detailed examination of bulk tunneling, which has not been done.  We shall, as in [\onlinecite{FFN}], adopt it as part of the definition of the tunneling operator, and adapt it to our case of a double point contact.

\begin{figure}[tbh!]
\includegraphics[width=2in]{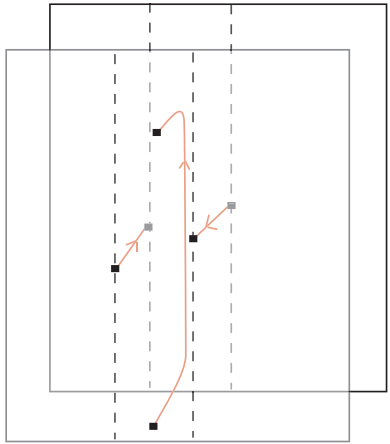}
\caption{Tunneling geometry with a test quasihole between the two point contacts.  The test quasihole is created and annihilated on the same side of the hall bar}
\label{f1}
\end{figure}

We wish to probe the effect of test quasiholes, residing on anti-dots in the region between the two point contacts, on the tunneling current.  The main conceptual point we wish to stress is that to do the computation correctly, one must incorporate the creation and annihilation of these test quasiholes in the far past and future.  That is, we must include two more $\sigma_j$ operators - we will take them to be on the same side of the hall bar, creating the quasihole in the far past and annihilating it in the far future (see fig (\ref{f1})).  For simplicity, we will analyze just two situations: one without any test quasiholes, and one with a single test quasihole, and for easy comparison we will actually create a test quasihole in both situations, but keep it far away from the region of interest in one case (so that it has trivial braiding).  For definiteness, we will for now analyze the following correlator, one of two that contribute to the interference term:

\begin{equation}
\langle \sigma_2 (x_0, t_0) \sigma_2 (x_1, t) \sigma_1 (D- x_1, t) \sigma_2 (D-x_2,t') \sigma_1(x_2,t') \sigma_1 (x_0, -t_0) \rangle
\end{equation} Here the test quasihole is created at $(x_0, -t_0)$, and annihilated at $(x_0,t_0)$.  $t_0$ and $D$ are to be taken to infinity.  We find it convenient to send $D$ to infinity first, so the thing splits into a $4$ point and a $2$ point function by cluster decomposition.  Let's see what happens when $t_0$ goes to infinity.  Then $\sigma_2 (x_1,t)$ and $\sigma_1 (x_2,t')$ can be considered to be very close (in comparison with $\sigma_2 (x_0, t_0)$ and $\sigma_1 (x_0, -t_0)$), so we can use the operator product expansion on them.  Clearly the leading contribution comes from the identity, so that's the only one we need to retain in the limit.  The $4$-point function is then just the $2$-point function of $\sigma_2 (x_1,t)$ and $\sigma_1 (x_2,t')$ times an overall constant having to do with the preparation and annihilation of the test quasihole.

\begin{figure}[tbh!]
\includegraphics[width=2in]{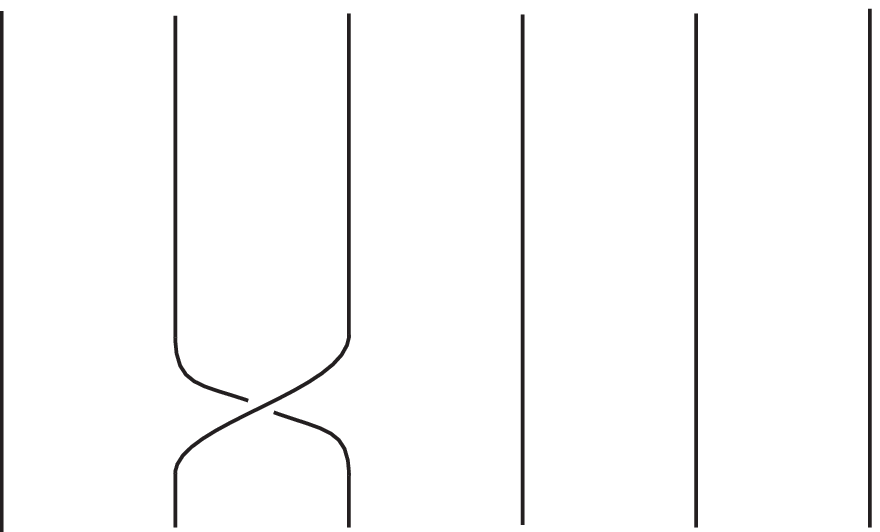}
\caption{Projection of the quasihole worldlines in the case of a test quasihole braiding nontrivially with the others}
\label{fb1}
\end{figure}

\begin{figure}[tbh!]
\includegraphics[width=2in]{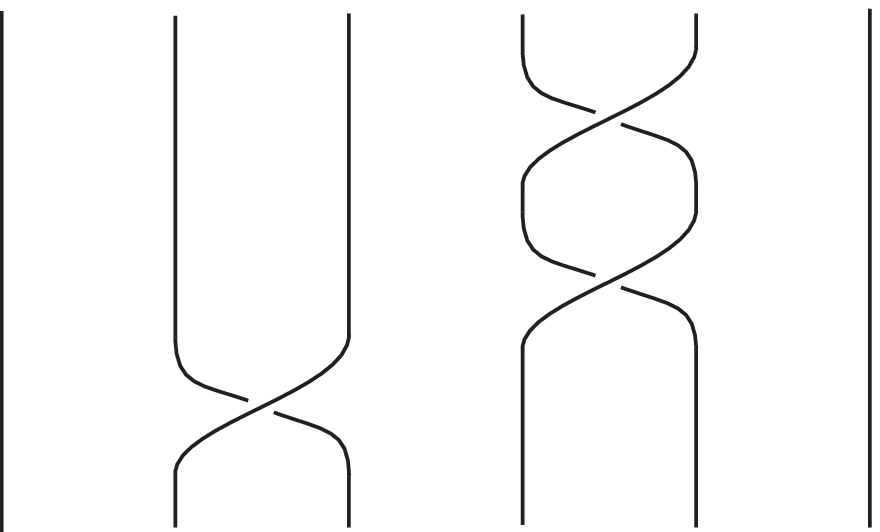}
\caption{Projection of the quasihole worldlines, together with a test quasihole braiding trivially}
\label{fb2}
\end{figure}

From the above paragraph, we see that it is very convenient to work in the basis where the fusion channels of the pairs $(\sigma_2(x_0,t_0), \sigma_1(x_0, -t_0))$, $(\sigma_2(x_1,t), \sigma_1(x_2,t'))$, and $(\sigma_1 (D-x_1,t), \sigma_2 (D-x-2,t'))$ are known, because only in the case when all those fusion channels are the identity do we get a nonzero contribution.  But the conformal block we're interested in is specified in a different basis.  To express it in the new basis, we proceed as in [\onlinecite{FFN}] and project the worldlines of the quasiholes (including the test quasihole) to two dimensions; the linear transformation necessary to enact the basis change is the matrix representing the resulting braid.  

The conformal block we start with is:

\begin{equation} 
\begin{picture}(210,30) 
\put(-70,3){$I$} 
\put(-65,20){$(x_2,t')$} 
\put(-30,20){$(D-x_2,t')$} 
\put(27,20){$(x_1,t)$} 
\put(65,20){$(x_2,t')$}
\put(112,20){$(x_0,-t_0)$}
\put (157,20){$(x_0,t_0)$}
\put(15,3){$I$}
\put(105,3){$I$} 
\put(-70,0){\line(1,0){285}} 
\put(-50,0){\line(0,1){15}} 
\put(-10,0){\line(0,1){15}} 
\put(40,0){\line(0,1){15}} 
\put(80,0){\line(0,1){15}} 
\put(130,0){\line(0,1){15}}
\put(170,0){\line(0,1){15}}
\put(195,3){$I$} 
\end{picture} 
\label{cb1} 
\end{equation} We want to expand it in the new basis above, and in particular get the coefficient of
\begin{equation} 
\begin{picture}(210,30) 
\put(-70,3){$I$} 
\put(-65,20){$(x_2,t')$} 
\put(-30,20){$(x_1,t)$} 
\put(12,20){$(D-x_2,t')$} 
\put(67,20){$(x_2,t')$}
\put(112,20){$(x_0,-t_0)$}
\put (157,20){$(x_0,t_0)$}
\put(15,3){$I$}
\put(105,3){$I$} 
\put(-70,0){\line(1,0){285}} 
\put(-50,0){\line(0,1){15}} 
\put(-10,0){\line(0,1){15}} 
\put(40,0){\line(0,1){15}} 
\put(80,0){\line(0,1){15}} 
\put(130,0){\line(0,1){15}}
\put(170,0){\line(0,1){15}}
\put(195,3){$I$} 
\end{picture} 
\label{cb2} 
\end{equation}

Depending on whether or not the test quasihole resides in the region between the two point contacts, we have two possible braiding histories, shown in figures (\ref{fb1}) and (\ref{fb2}).  Because, due to the considerations outlined below, we only care about the coefficients up to phase, we only need to determine the matrix for the linear transformation between the two bases up to phase.  Thus, even though a priori we would have to deal with the Moore-Seiberg braid matrix for the parafermions (and the associated $U(1)$ theory), we really only need the braid matrix in the associated quantum group picture.  Here, both $\sigma_1$ and $\sigma_2$ correspond to the $2$-dimensional spin $1/2$ representation of $U_q(\text{sl}_2)$, $q=\exp (2 \pi i / 5)$.  Labeling conformal blocks using the Bratteli diagram notation, we see that we start off with the block represented by the path $[(0,0), (1,1), (2,0), (3,1), (4,0), (5,1), (6,0)]$, which for convenience we will designate via  shortened notation by $[0101010]$.  We see that when the braids in (\ref{fb1}) and (\ref{fb2}) act, everything happens in the subspace spanned by $[01x1y10]$, $x,y=0,2$, and each transposition acts either only on $x$ or only on $y$ (they commute).  The relevant unitary matrix enacting the transposition is [\onlinecite{JS}]:

\begin{equation}
U_{i,i+1} = \frac{e^{-\pi i / 10}}{\tau} \begin{pmatrix}
- e^{-2 \pi i / 5} & - \sqrt{\tau} \\
-\sqrt{\tau} & e^{2 \pi i /5}
\end{pmatrix}
\end{equation}where $i=2,4$.  The $2$-element basis in which we write the matrix $U_{2,3}$ is $x=2,0$, and the one corresponding to $U_{4,5}$ is $y=2,0$.  Note also that, using some identities involving the golden mean, we have:

\begin{equation}
{U^2}_{4,5} = \frac{e^{-\pi i / 5}}{\tau^2} \begin{pmatrix}
- e^{4 \pi i / 5} & i \sqrt{ \tau (\tau+2)} \\
i \sqrt{ \tau (\tau+2)} & -e^{-4 \pi i /5}
\end{pmatrix}
\end{equation}We have

\begin{eqnarray}
\label{feqn}
\langle [0101010] | U_{2,3} | [0101010] \rangle = \frac{1}{\tau} \\ \label{feqn1}
\langle [0101010] | U_{2,3} {U^2}_{4,5} | [0101010] \rangle = \frac{1}{\tau^3}
\end{eqnarray}

Let us now use this to obtain the $C_{jk}$.  First, let's see what we can surmise on general grounds.  By tuning $\Gamma_2$ to $0$ and using the fact that $\langle j(t) \rangle$ is real, we find that $C_{11}$ is real.  Also, by symmetry, $C_{11} = C_{22} = C$.  Again because the current is real we see that $C_{21} = {C_{12}}^*, C_{12} = |C_{12}| e^{i \theta}$.  The phase $\theta$ can in principle be determined, but we ignored it in the calculations above.  This is because the effect of changing $\theta$ is the same as that of changing the phases of $\Gamma_1$ and $\Gamma_2$, which are hard to determine anyway and are sensitive to the Aharonov-Bohm effect.  

The results in (\ref{feqn}) and (\ref{feqn1}) show that when a quasihole is present, $|C_{12}|$ is diminished by a factor of $1 / \tau^2$.  Similar topological calculations (with only the spatial coordinates of the points involved changing) show that the other constants, $C_{11}$ and $C_{12}$, are always equal, and equal to the $|C_{12}|$ in the case when there is no test quasihole in the region between the point contacts.  Thus we have

\begin{equation}
{\langle j(t) \rangle}_{\text{no qh.}} = A {| \omega_J |}^{-3/5} \text{sgn} (\omega_J) {v_n}^{-4/15} {v_c}^{-2/15} \left({|\Gamma_1|}^2 + {|\Gamma_2|}^2 + (\Gamma_1 {\Gamma_2}^* + {\Gamma_1}^* {\Gamma_2}) H(\omega_J {|{x_1 - x_2}|} / v_n, v_c / v_n) \right)
\end{equation}

\begin{equation}
{\langle j(t) \rangle}_{\text{qh.}} = A {| \omega_J |}^{-3/5} \text{sgn} (\omega_J) {v_n}^{-4/15} {v_c}^{-2/15} \left({|\Gamma_1|}^2 + {|\Gamma_2|}^2 + {\frac{1}{\tau^2}}(\Gamma_1 {\Gamma_2}^* + {\Gamma_1}^* {\Gamma_2}) H(\omega_J {|{x_1 - x_2}|} / v_n, v_c / v_n) \right)
\end{equation} where $A$ is an unimportant constant.

We thus observe that even without looking at phases, nonabelian statistics have a robust effect of changing the magnitude of the coefficient of the interference term in the tunneling current.  This is in line with earlier arguments of [\onlinecite{BSS}] for the differential conductivity.

\section{Acknowledgments}
I would like to thank Eddy Ardonne, Michael Freedman for discussions, and Chetan Nayak for discussions and a careful reading of the draft.  This research was supported partly by the NSF under grant no. -0244728, the SITP, and Microsoft.

\end{document}